\documentclass[12pt]{article}
\usepackage[utf8]{inputenc}
\usepackage{authblk}
\usepackage{setspace}
\usepackage[margin=1.25in]{geometry}
\usepackage{graphicx}
\usepackage{subcaption}
\usepackage{amsmath}
\usepackage{float}
\newcommand{\p}{\ensuremath{\partial{}}}

\usepackage[backend=bibtex,style=nejm, 
citestyle=numeric-comp,
sorting=none]{biblatex}

\begin{document}
\title{Realizing late-time cosmology in the context of Dynamical Stability Approach}
\author[1*]{Anirban Chatterjee}
\author[1]{Saddam Hussain}
\author[1]{Kaushik Bhattacharya}
\affil[1]{Department of Physics, Indian Institute of Technology Kanpur, Kanpur 208016, India}
\affil[*]{Address correspondence to: anirbanc@iitk.ac.in}

\onehalfspacing
\maketitle

\date{}

\begin{abstract}

We examine the scenario of non-minimally coupled relativistic fluid and $k$-essence scalar field in a flat Friedmann-Lemaitre-Robertson-Walker universe.  By adding a non-minimal coupling term in the Lagrangian level, we study the variation of Lagrangian with respect to independent variables, which produces modified scalar field and Friedmann equations.  Using dynamical stability approach in different types of interaction models with two types of scalar field potential, we explore this coupled framework. Implementing detailed analysis, we can conclude our models can able to produce stable late-time cosmic acceleration. 

\end{abstract}

\vspace{-1.2cm}

\section{Introduction}
Standard model of cosmology ($\Lambda$-CDM model) \cite{Turner:1992}, mainly suffers from two drawbacks, first one is the fine-tuning problem and second one is a cosmic-coincidence problem. In this standard model of cosmology, $\Lambda$ represents the cosmological constant and CDM denotes the cold-dark matter. Another important downside of the $\Lambda$-CDM model from the observational perspective is the discrepancy between the local measurement of present observed value of Hubble’s constant and the value predicted from the Planck experiment using $\Lambda$-CDM model \cite{DiValentino:2021izs}.  These fundamental discrepancies motivate us to study different kinds of cosmological models found in the non-minimally coupled field-fluid sectors \cite{Chatterjee:2021ijw},\cite{Hussain:2022osn},\cite{Bhattacharya:2022wzu}. Based on these above considerations, we build a theoretical framework for a coupled field-fluid sector, where field sector is made of a non-canonical scalar field ($k$-essence sector \cite{Armendariz-Picon:1999hyi},\cite{Armendariz-Picon:2000ulo}) and the fluid sector is composed of pressure-less dust. The non-minimal coupling term is introduced at the Lagrangian level. We employ the variational approach \cite{Brown:1992kc} with respect to independent variables that produce modified $k$-essence scalar field equations and the Friedmann equations. Then we analyze this coupled field-fluid framework explicitly using the dynamical system technique \cite{Bahamonde:2017ize}, considering two forms of the scalar field potential, \textit{viz.}  inverse power-law type \cite{Chatterjee:2021ijw} and constant type \cite{Hussain:2022osn}. After examining these scenarios, both models can produce accelerating attractor solutions and satisfy adiabatic sound speed conditions.
\vspace{-1.2cm}
\section{Theoretical Framework}
Total action for this non-minimally coupled field-fluid sector \cite{Chatterjee:2021ijw} can be written as,
\begin{eqnarray}
S&=&\int_\Omega d^4x \left[\sqrt{-g}\frac{R}{2\kappa^2}-\sqrt{-g}\rho(n,s) +
J^\mu(\varphi_{,\mu} + s\theta_{,\mu} + \beta_A\alpha^A_{,\mu})    
-\sqrt{-g}{\mathcal L}(\phi,X)\right.\nonumber\\
& &\left.-\sqrt{-g}f(n,s,\phi,X)\right]\nonumber,\quad \quad \mbox{(Here, $\kappa^2=8\pi G$)}
\label{act}
\end{eqnarray}
\vspace{-1.2cm}

The first term corresponds to the action's gravitational part, and the second and third terms represent the action related to the relativistic fluid sector. Fourth term implies the action for the $k$-essence scalar field. Finally, the last term describes the action for the non-minimal coupling. Non-minimal coupling term ($n,s,\phi,X$) depends on the functions of both sectors of field ($\phi, X$) and fluid ($n,s$). Varying the grand action respect to $g_{\mu \nu}$, we get the total energy-momentum tensor for this coupled system $T_{\mu \nu}^{\rm tot.}=T_{\mu \nu}^{(\phi)} + T_{\mu \nu}^{(M)} + T_{\mu \nu}^{({\rm int})}$. Total energy-momentum tensor has been conserved, but the individual components have not been conserved. Variation of above action with respect to the independent variables produce the modified field equation as, ${\mathcal L}_{,\phi} + \nabla_\mu ({\mathcal L}_{,X} \nabla^\mu \phi) + f_{,\phi} + \nabla_\mu (f_{,X} \nabla^\mu \phi)=0$.  In the context of a flat-FLRW metric ($ds^2 = -dt^2 +a(t)^2 d{\bf x}^2$), the above equation can be redundant for both potential and kinetic terms dependent scalar field case \cite{Chatterjee:2021ijw} as, $\left[ \mathcal{L}_{,\phi} + f_{,\phi}\right] - 3H\dot{\phi}\left[  \mathcal{L}_{,X} + f_{,X} \right] +  \dfrac{\p }{\p X} (P_{\rm int} +f ) (3H \dot{ \phi})\, -
\ddot{ \phi} \left[(\mathcal{L}_{,X} + f_{,X}) + 2X (\mathcal{L}_{,XX} + f_{,XX}) \right] -\dot{ \phi}^2 (\mathcal{L}_{,\phi X} + f_{,\phi X }) = 0$. And for purely kinetic dependent scalar field \cite{Hussain:2022osn} as, $	- 3H\dot{\phi}\left[  \mathcal{L}_{,X} + f_{,X} \right] +  \dfrac{\p }{\p X} (P_{\rm int} +f ) (3H \dot{ \phi})\, - 
	\ddot{ \phi} \left[(\mathcal{L}_{,X} + f_{,X}) + 2X (\mathcal{L}_{,XX} + f_{,XX}) \right]   = 0$. \\
\vspace{-1.2cm}
\section{Results \& Discussion}
Utilizing some dimensionless variables, we recast field, fluid, and interaction sectors in terms of them. Friedmann equations are also modified in terms of these variables and acted like constraint equations of the dynamical system. Depending on the total number of used independent variables dimension of the phase space has been decided. Depending on the dimension of the phase space, we have divided our analysis into two cases. 

\begin{itemize}

\item  \textbf{I. Algebraic Coupling with arbitrary scalar field potential:} To study this case, we can choose the form of interaction \cite{Chatterjee:2021ijw}, $f = \rho \alpha \left(\dfrac{\phi}{\kappa}\right)^m \beta X^n$ ($\alpha, \beta, m, n$ are constant). Using linear stability approach, we get a total of eight critical points from the phase space analysis, in which two sets are stable and other two sets are showing saddle type nature. From Evolutionary dynamics of this coupled system suggest that a late-time stable accelerating phase can be achieved through this non-minimal coupling. Transfer of energy from field-to-fluid and finally fluid-to-field can also be observed.  Total EOS parameter of this coupled sector saturates near $-1$ at late-time era.  At present epoch energy density of dark matter and dark energy are in same order of magnitude, also observed from here.
\vspace{-0.5cm}
\begin{figure}[H]
\begin{minipage}[b]{0.5\linewidth}
\centering
\includegraphics[scale=.32]{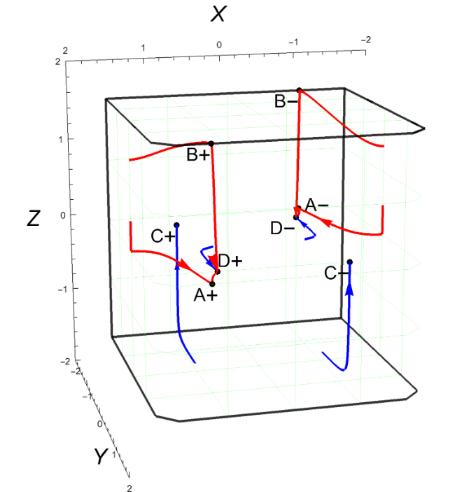} 
\subcaption{Phase space plot for $n=1, m=3,  \delta = 10$ } 
\label{fig:Ma1}
\end{minipage}
\hspace{0.2cm}
\begin{minipage}[b]{0.5\linewidth}
\centering
\includegraphics[scale=.25]{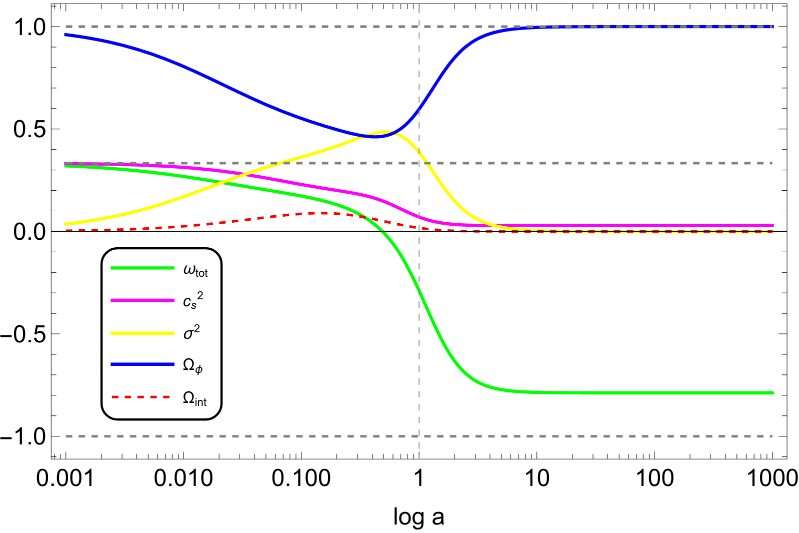}
\subcaption{Evolution plot for the model parameters $n=1,  m=3, \delta = 10$}
\label{fig:Ma2}
\end{minipage}
\caption{Phase space and evolution plots for Case-I (For details see \cite{Chatterjee:2021ijw})}
\end{figure}


\item  \textbf{II.Algebraic  Coupling with constant scalar field potential:} The form of interaction for this case \cite{Hussain:2022osn} has been chosen $f= g V_0 \rho^{q} X^\beta M^{-4q}$ (where, $q=-1, g, V_0, \beta$ are constant). Due to the absence of the potential term the dimension of the phase space is reduced into 2-D. Utilizing the phase space analysis of dynamical system, we have also found one stable critical for this type of system. Phase space is constrained by the modified Friedmann equation, accelerating universe, and sound speed condition. Evolution plot suggests that late-time stable accelerating phase can be achieved and an energy transfer between field to fluid sector has also been observed through this framework of non-minimal coupling. Total EOS parameter of this coupled sector saturates near $-1$ at late-time era. 
\vspace{-0.48cm}
\begin{figure}[H]
\begin{minipage}[b]{0.5\linewidth}
\centering
\includegraphics[scale=.18]{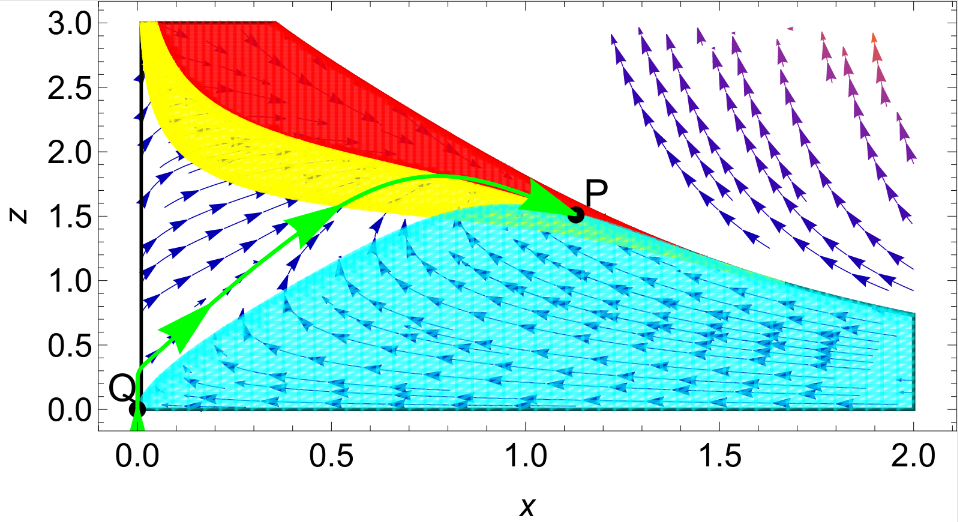} 
\subcaption{Phase space plot for $ A = 1/2, B=1/3, q=-1,\alpha =1, g=1/2,\beta=1/3,  \mathcal{M} = 1$ } 
\label{fig:Mb1}
\end{minipage}
\hspace{0.2cm}
\begin{minipage}[b]{0.5\linewidth}
\centering
\includegraphics[scale=.5]{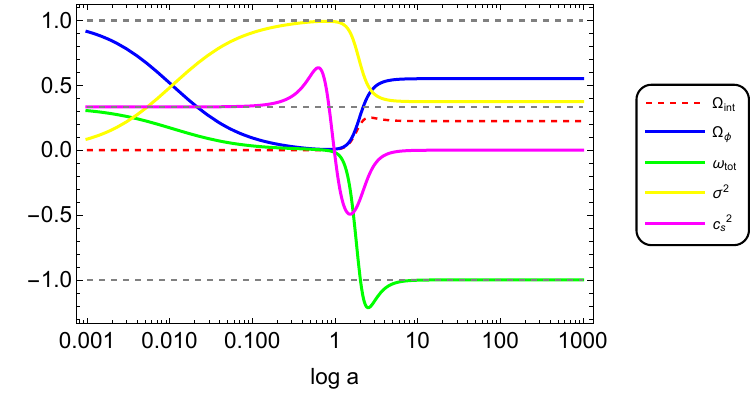}
\subcaption{Evolution plot for the model parameters $ A = 1/2, B=1/3, q=-1,\alpha =1, g=1/2,\beta=1/3,  \mathcal{M} = 1$}
\label{fig:Mb2}
\end{minipage}
\caption{Phase space and evolution plots for Case-II (For details see\cite{Hussain:2022osn})}
\end{figure}

\end{itemize}

\vspace{-1cm}

\printbibliography

\end{document}